# Multipacting

*R. Parodi*
INFN-Genova, Genoa, Italy

**Abstract**
Multipacting (MP) is a resonant electron discharge, often plaguing radiofrequency structures, produced by the synchronization of emitted electrons with the RF fields and by the electron multiplication at the impact point with the surface of the structure. The current of re-emitted electrons grows via true secondary re-emission when the secondary yield for the primary electron impact energy is greater than one. A simple example (MP in short-gap accelerating axial-symmetric cavities) allows an analytical solution of the equation of motion, giving both the synchronization (kinematics) and multiplication (impact energy) conditions as a function of the gap voltage (or accelerating field). Starting from this example a thorough discussion of MP discharges in axial-symmetric accelerating structures will be given and some poor man's rules are given to estimate the critical cavity field levels to meet the kinematic condition for resonance. The results of these poor man's rules are compared with computer simulations of MP discharges obtained by a statistical analysis of the re-emission yield for impinging electrons versus RF field level in the accelerating structure.

## 1 Introduction

Multipacting (MP) is a resonant RF electron discharge in vacuum with electron multiplication due to the secondary electron re-emission process. The discharge is mainly encountered in RF accelerating structures where the combination of RF fields and clean surfaces of high secondary yield metals like copper, aluminium, or niobium will enhance the electron multiplication of the electrons at each impact.

This kind of discharge has also been encountered in RF tubes and waveguide devices since the pioneering times of radio broadcasting and microwave radar development. This phenomenon is often also encountered at the interface between vacuum and dielectrics such as RF windows in waveguides or ceramic feed-through in power devices such as RF couplers for accelerating cavities or broadcasting power tubes and devices.

Multipacting was a real head ache in the early 1970s for the pioneering development of superconducting accelerators at Stanford University, both for SLAC and HEPL.

In this application the combination of high quality factors giving high field at low power, clean niobium surfaces with high secondary yields, and superconductivity were ideal conditions to create very strong and quite insurmountable MP barriers.

## 2 Basic facts about multipacting

To have multipacting you need the occurrence of two conditions:

1. electron synchronization with the RF fields

2. electron multiplication via secondary electron re-emission.

Both conditions need to be met by the electron clouds trapped in the RF fields, to produce a real MP barrier.

The first is the kinematics of the electrons flying inside the RF structures, and is met when the time between two electron impacts is an integer number of half RF periods.

In this way the starting conditions for the electrons re-emitted at each structure are always the same and the motion of the re-emitted electrons at the given field always follows the same trajectories, and the process repeats itself indefinitely.

The second condition sets the mechanism giving a runaway discharge leading possibly to the total depletion of the energy stored in the cavity, and sometimes setting a limitation to the accelerating field already achievable in a reliable way in an accelerating structure.

To have electron multiplication we need the secondary emission coefficient $\delta$ greater than one. (Even slightly.)

Owing to the cyclic properties of the motion stated by the first condition, the multiplication factor $Y$ after $N$ impacts has an exponential growth given by

$$Y = \delta^N \qquad (1)$$

where $\delta$ is the secondary emission coefficient and $N$ the number of impacts. With the time between impacts ~ of the RF period by condition 1 (RF synchronization) the number of impacts per second is ~ the RF frequency, ranging from $10^8$ to $10^9$: assuming a secondary emission coefficient ~ 1.1 we get, after 100 nanoseconds, a multiplication factor ~$10^4$, more than enough to give serious problems to the cavity behaviour in a time short enough to make impossible any attempt to intervene actively to compensate the additional electron loading.

To have secondary electron multiplication the impact energy $U$ of the primary electron for metals is usually in the range 50 eV to 1500 eV, see Ref. [1]

The starting condition for a re-emitted secondary electron is a kinetic energy ~ 2 eV and re-emission direction randomly distributed like $\cos\theta$ around the normal direction to the re-emitting surface at the impact point.

The energy and re-emission angle for the secondary electrons allow us (in the first approximation) to neglect the effect on the starting condition in a first-order estimate of the electron motion.

Figure 1 is a plot of the secondary emission electron yield for a metal as a function of the primary electron impact energy from 10 eV to 10 000 eV.

Good metallic conductors with low resistive losses used in RF accelerating structures and devices (both normal conductors such as copper and silver, and superconductors such as niobium) have a secondary emission coefficient $\delta_o$ ranging from 1.5 to 2.5 depending upon the microscopic status of the surface, the oxidation status, and the surface contamination.

Usually oxide layers (with good dielectrics) have quite high $\delta_o$. Carbides and nitrides have low $\delta_o$.

Titanium and stainless steel have $\delta_o$ lower than one but, unfortunately, have high resistive losses and are not suitable for high-power applications. Despite the quite high losses, stainless steel is often used in very special cases such as the short-gap buncher cavities of RF injectors operating in the magnetic field of a guiding solenoid used to keep focused the low-energy beam coming from the DC electron gun to increase the capture efficiency of the injector. In this case the need for a somewhat higher RF power in operation is more than compensated by the reliable operation of the injector free from any resonant discharge.

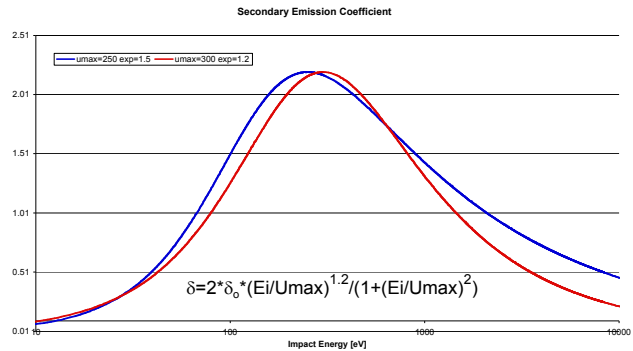

**Fig. 1:** Secondary emission yield as a function of the primary electron impact energy for a metal with maximum yield 2.2 at two different values of the primary electron impact energy (250 eV blue line, and 300 eV red line)

## 3 Resonant discharges in axial-symmetric cavities

The two conditions needed to have a resonant TF discharge are valid in general, but give no way to predict or to diagnose resonant discharge effects in accelerating cavities and RF structures.

In the following sections we will show some examples of MP discharges that can be analytically solved as two-point MP in short-gap-length axial-symmetric cavities with gap length $\ll \lambda$ RF, or at least we can give a sort of rule of thumb to estimate the field where the kinematic of the emitted electrons should produce synchronization with the RF fields as in the one-point MP discharges at the equatorial region of axial-symmetric accelerating cavities.

### 3.1 Two-point MP discharges in short-gap cavities $L \ll \lambda$

This discharge happens when the re-emitted electrons strike cavity regions with field of opposite sign, as an example flying across a short-gap.

This condition is usually fulfilled in short-gap cavities as bunchers and catchers following the DC electron guns at the linac injectors. A typical cross section of a buncher cavity is shown in Fig. 2, the plot gives the shape of the cavity with the electric-field lines superimposed.

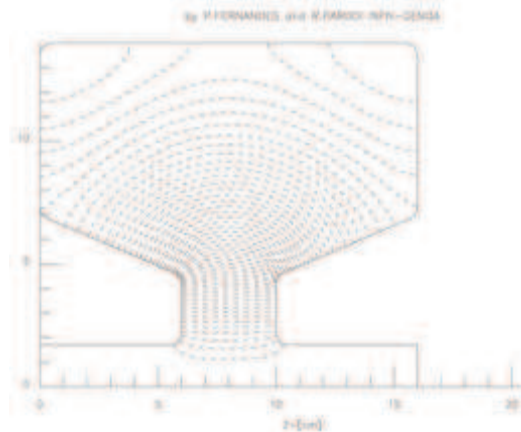

**Fig. 2:** Short-gap cavity developed at LAL Orsay as sub harmonic buncher for the linac injectors

The semi-lumped cavities such as the ones used in the old time storage rings like ADONE [2] are quite prone to this type of MP discharges in the gap being substantially a resonant coaxial line (the inductor) heavily loaded by a parallel plate capacitor (Fig. 3).

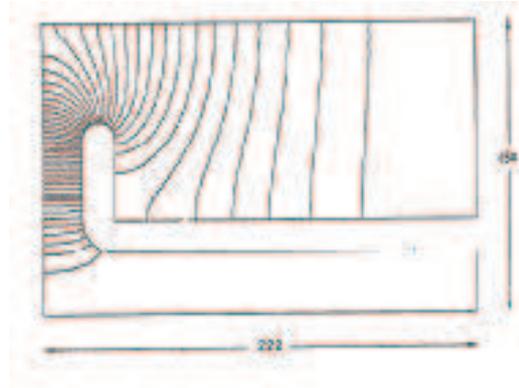

**Fig. 3:** Plot of 51.4 MHz semi-lumped ADONE cavity. The plot shows a half cavity with superimposed electric field lines.

From inspection of the cavity field distribution it easily seen that the field distribution in the gap region is very similar to the field distribution in a parallel plate capacitor of length $L$ (Fig. 4). The field variation is sinusoidal in time with angular frequency $\omega$ and amplitude $E_o$ V.

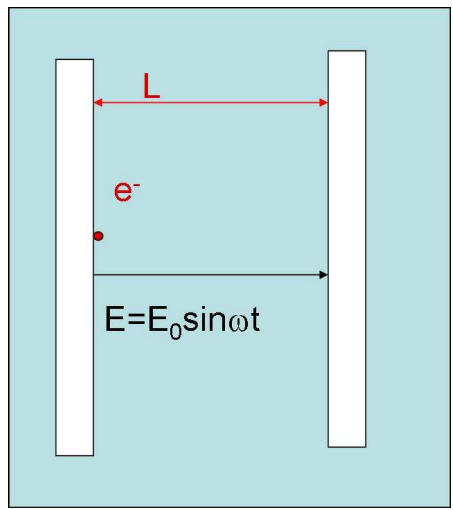

**Fig. 4:** Parallel plate capacitor model of the short-gap cavity like the ones shown in Figs. 2 and 3. The model is used to derive the approximate solution leading to the MP barrier values for the given gap length $L$ and cavity operating frequency $f$.

Under these assumptions we can write the equation of motion for an electron starting from one of the capacitor plates.

$$\ddot{x} = \frac{eE_o}{m_o} \sin \omega t .  \qquad (2)$$

Assuming as starting condition an electron emitted with zero energy when the field is zero (a good approximation of the 2 eV re-emission energy for true secondary electrons), the integration of the equation of motion is straightforward and leads to

$$\dot{x} = \frac{eE_o}{m_o\omega}(1 - \cos\omega t), \quad (3)$$

$$x = \frac{eE_o}{m_o\omega}t + \frac{eE_o}{m_o\omega^2}\sin\omega t, \quad (4)$$

using the starting condition re-emission energy equal to zero and starting point at $x = 0$.

By imposing the boundary condition of electric field reversal at the impact on the plate at position $x = L$, we get the synchronization condition for the re-emitted electron impacting the opposite plate after an **odd integer number** of **half RF periods**

$$t = \frac{T}{2}(2n-1) = \frac{1}{2\nu}(2n-1) = \frac{\lambda}{2c}(2n-1), \quad (5)$$

where $T$ is the RF period, $\nu$ is the RF frequency, $\lambda$ is the RF wavelength, and $n$ is an integer number giving the MP discharge order.

Equations (4) and (5) combined together give an expression for the electric field value (or the gap voltage) corresponding to the $n$-th barrier in terms only of the gap $L$, the RF wavelength $\lambda$, and the electron rest mass $m_o$ given in Eq. (6) and for the impact energy given in Eq. (7)

$$V = L \cdot E_0 = 4\pi \frac{m_0 c^2}{e}\left(\frac{L}{\lambda}\right)^2 \frac{1}{(2n-1)} \quad (6)$$

$$U = 8m_0 c^2\left(\frac{L}{\lambda}\right)^2 \frac{1}{(2n-1)^2}. \quad (7)$$

In this way we can forecast the possible MP barrier in short-gap cavities knowing only the gap length and the operating frequency, and even the order of the barrier just using a simple Excel spreadsheet like the one reported in Table 1.

**Table 1:** First 10 multipacting levels for a short-gap cavity $L = 4$ cm operating at 500 MHz ($L \ll \lambda$)

| Barrier order n | Gap voltage [V] | Impact energy [eV] | $E_0$ [V/m] |
|---|---|---|---|
| 1 | 2.85E+04 | 1.82E+04 | 713.5E+3 |
| 2 | 9.51E+03 | 2.02E+03 | 237.8E+3 |
| 3 | 5.71E+03 | ***7.27E+02*** | 142.7E+3 |
| 4 | 4.08E+03 | ***3.71E+02*** | 101.9E+3 |
| 5 | 3.17E+03 | ***2.24E+02*** | 79.3E+3 |
| 6 | 2.59E+03 | ***1.50E+02*** | 64.9E+3 |
| 7 | 2.20E+03 | ***1.08E+02*** | 54.9E+3 |
| 8 | 1.90E+03 | 8.08E+01 | 47.6E+3 |
| 9 | 1.68E+03 | 6.29E+01 | 42.0E+3 |
| 10 | 1.50E+03 | 5.03E+01 | 37.6E+3 |

The potentially dangerous barriers are highlighted in red bold italic in Table 1. The primary electrons for these barriers have impact energy in the range 100–1500 eV, corresponding to a secondary emission coefficient greater than one in copper or aluminium.

The comparison with the MP barriers, as measured in the built cavities and computed by computer simulations of the discharge, is in extremely good agreement with the barrier found by the analytical model.

A typical plot of the trajectory simulation in the cavity is shown on Fig. 5.

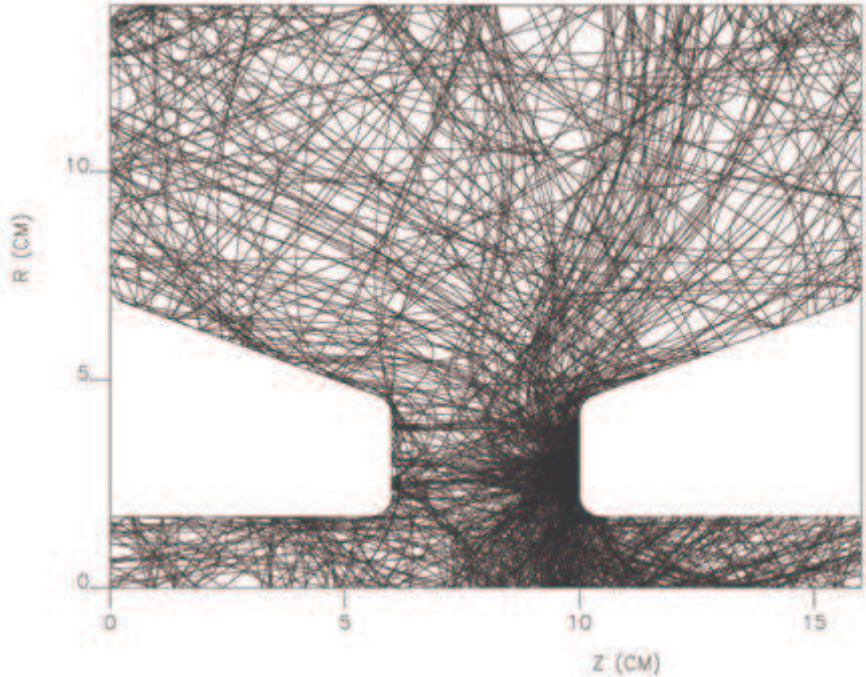

**Fig. 5:** Plot of the trajectories in the short-gap cavity having two-point MP discharges in the gap. Despite the spread of the secondary electrons all over the cavity, only the electrons travelling across the gap survive 5000 impacts with a yield greater than one.

### 3.2 One-point MP at cavity equator of axial-symmetric cavities

This kind of discharge can occur when electrons impact the cavity surface almost at the same place. The synchronization is obtained when the time of flight is about equal to an integer number of RF periods; in this way the re-emitted electrons have the very same starting condition as the primary (position and RF field).

Again multiplication is achieved if the impact energy of the primary is in the range 50–1500 eV to have the secondary yield greater than one.

In axial-symmetric accelerating $TM_{01}$ cavities this resonant discharge happens at the equatorial region where the magnetic field is high (and the electric field is quite low).

In a pill-box $TM_{010}$ cavity at the equator, the electric field is zero (due to the boundary condition) and the magnetic field is roughly 90% of the maximum value; in a practical cavity with beam openings, a small component radial electric field is produced at the cavity outer wall and this field can, in some cases, give to the re-emitted electrons the right amount of energy to produce true secondary electrons with re-emission yield greater than one.

For this kind of MP discharge no simplified analytical model is possible, the energy gain of the electron around the trajectory being strongly dependent on the detail of the field distribution; a thorough analysis of the MP was performed through computer simulations computing the electron trajectories and evaluating the cumulative yield after a pre-fixed number of impacts.

Nevertheless a poor man's estimation can possibly help the cavity designers to guess the field levels where the occurrence of MP discharges should more probably develop.

Let us consider the motion of an electron close to the equatorial region of a pill box where the RF magnetic field is near to the maximum and the electric RF field is close to zero.

The trajectories are roughly cyclotron orbits and the synchronous motion condition 'one impact per integer number of RF periods $T$' gives the approximated expression of Eq. (8).

$$\frac{1}{T \cdot n} = \frac{\upsilon}{n} = \frac{eB_0}{2\pi m_0}. \tag{8}$$

Solving for the field value we get, for the possible value of the local magnetic field at the *n*-th MP level, Eq. (9) with ν the RF frequency, *e* the electron charge, and $m_o$ the electron rest mass.

$$B_0 = \frac{2\pi \upsilon m_0}{e} \frac{1}{n}. \tag{9}$$

A typical plot of the trajectories for the MP barriers at the equator is shown on Fig. 6

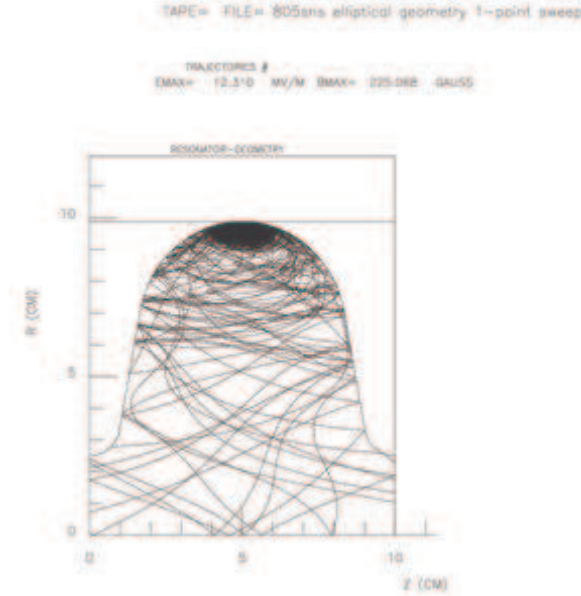

**Fig. 6:** Typical trajectories for the first-order MP barrier at the equator as computed by numerical integration of the equations of motion of a re-emitted electron in the RF field. The operating frequency is 805 MHz, the resonant condition is found at the magnetic field level of 22.5 mT.

Using for the local magnetic field the rms value, to account for the sinusoidal time variation of the RF fields, we get a $B_o$ value of ~ 28 mT/GHz as resonant field value for the MP discharge with $n = 1$.

In the reported example the cavity frequency is 805 MHz, the resonant field is 22.5 mT as foreseen by the semi-empirical formula, the total secondary yield after 200 impacts is ~$10^{-2}$ giving insufficient secondary electrons to sustain a discharge, even if synchronization with the RF field is achieved.

The discharge is quenched because of the focusing of the re-emitted electron at the cavity equator due to the strong component of the electric field in the axial direction.

At this position the electric field is zero by symmetry, and the energy gain is not enough to give a significant secondary multiplication.

Furthermore at the cavity equator where the electric field is zero, the starting condition for the secondary electrons starts to be relevant.

Even if the re-emission energy of 2 eV gives to the secondary electrons a very small speed, the random re-emission angle distribution (like cosθ around the normal direction) can produce trajectories landing in the opposite half-cavity, where the fields are reversed, and the electrons lose the synchronization.

### 3.3 Two-point MP at cavity equator of axial-symmetric cavities

Finally we need to mention that at the cavity equator a two-point MP discharge is still possible when secondary electrons emitted by the cavity surface near the equator are bent in a cyclotron-like trajectory by the local RF flying across the cavity midplane striking the cavity surface in a point nearly symmetric to the starting point. This phenomenon was first seen at CERN in the prototype cavities for the LEP project [3] and extensively analysed in Ref. [4].

The electrons bent by the RF magnetic field strike a cavity region with opposite electric field: as in the short-gap barriers, the only possibility left to get synchronization with the RF field and sustain the resonant discharge is to have the primary electron striking the surface after an odd integer number of half RF periods, the same condition of Eq. (5).

Again we can modify the poor man's rule used to guess the value of the magnetic field needed to match the resonant condition in the one-point MP at the equator of the accelerating cavity.

Starting from Eq. (10):

$$\frac{2}{T \cdot (2n-1)} = \frac{2\nu}{2n-1} = \frac{eB_0}{2\pi m_0}. \tag{10}$$

We get for the magnetic field the expression reported in Eq. (11):

$$B_0 = 2\frac{2\pi \nu m_0}{e}\frac{1}{(2n-1)}. \tag{11}$$

With the same meaning (and values) for $m_0$, $e$, and $\nu$ as Eq. (9) we get for the synchronous field of the MP level with $n = 1$, a value $B_o$ of 56 mT/GHz; Fig. 7 reports the trajectory distribution for this barrier as obtained by computer simulation.

In the reported case the field for the trajectory synchronization is 45 mT, and the total yield after 200 impacts is $10^4$ because the re-emitted electrons go through the cavity midplane landing in the half-cell opposite to the take-off point. On the way the field reversal (and the higher surface field values) allowed the re-emitted electrons to strike the cavity surface with enough energy to sustain the multiplication process.

A better understanding of the resonance effect is seen by reporting the number of electrons surviving a pre-fixed number of impacts as a function of the RF field intensity as obtained by the simulation of the electron trajectories in the RF fields of the cavity under analysis.

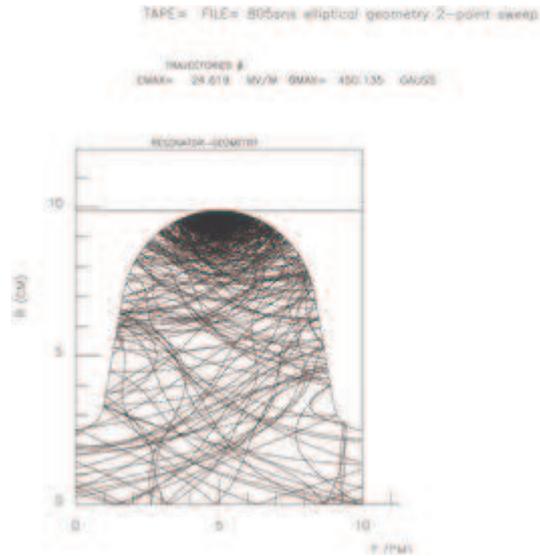

**Fig. 7:** Two-point MP barrier ($n = 1$) in the same cavity shown on Fig. 6. The operating frequency is 805 MHz, the resonant condition is found at the magnetic field level of 45 mT according to Eq. (11).

Figure 8 reports the number of electrons surviving a pre-fixed number of impacts, (800 in the given case).

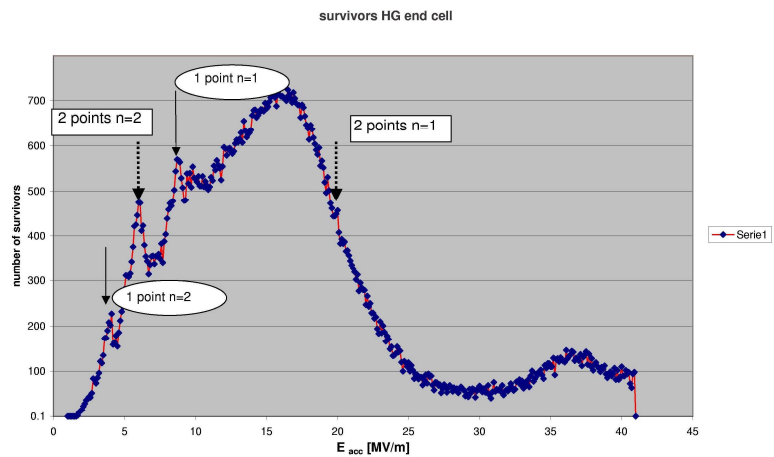

**Fig. 8:** Number of electrons surviving 800 impacts as a function of the accelerating field for the cavity geometry reported in Figs. 6 and 7. At the field level corresponding to the resonant condition for the MP barriers at the equator, the number of impacts shows a sharp peak in the number of surviving electrons.

It is evident that the number of 'surviving electrons' shows quite sharp peaks at the field level corresponding to the 'resonant field' obtained using Eqs. (10) and (11). This plot explains the method to be used to detect potential MP barriers in a given cavity geometry using the computer simulation of the emission process.

Given the cavity geometry, the analysis is performed in the following way:

1) Starting from the cavity geometry the RF field distribution is computed.
2) The field level is increased in steps from zero to the maximum field foreseen for the cavity operation.
3) At each field level electrons are started close to the possible MP region (usually the equator); the number should be large enough to give a reasonable statistical sample of surviving electrons.
4) The number, the yield, and mean impact energy of the surviving electrons are reported as a function of the cavity field levels.

The field levels of possible MP discharge are found as RF field intervals showing peaks in the number of electrons together with an enhancement in the mean yield of the secondary electrons when the impact energy of the electrons on the metallic surface is in the range giving a secondary emission coefficient greater than one (usually 100–1500 eV).

## 4 How to avoid MP discharges

From the previous discussions it appears clearly that there are only two ways to avoid or reduce the possibility of MP:

1) Reduce electron multiplication.
2) Avoid spatial focusing of electrons in critical regions to break time focusing and RF synchronization leading to resonant build-up of the secondary yield.

### 4.1 Reduce electron multiplication

A very radical way to avoid resonant discharge is to kill the electron multiplication using materials with secondary emission coefficient lower than one for the RF devices; in this way the number of re-emitted electrons decreases exponentially with the number of impacts, and the discharge is unable to self-sustain.

Unfortunately, good conductors used in RF applications such as copper, aluminium, or silver have secondary emission coefficient $\delta_0$ greater than one and strongly dependent on the surface oxidization and contamination status [5]; the same phenomenon happens, and is even worse, for the niobium in superconducting cavity applications. In oxidized niobium $\delta_0$ values up to 2.2 were measured [6].

Metals with $\delta_0$ lower than one, like titanium or stainless steel, usually have a poor electrical conductivity leading to high RF losses; for that reason the use of low secondary emission coefficient is very limited and used as a last resort when the application forces this choice; amorphous carbon or graphite have a very low secondary emission coefficient but are very good RF loads.

A typical application where stainless steel is used are short-gap buncher cavities operating in the magnetic field of a solenoid at the low-energy end of a linac injector as in the CLIO injector sub-harmonic buncher at 500 MHz [2].

Coating with film of low $\delta_0$ metals like titanium, using a film of thickness lower than the skin depth of the RF fields, avoids a large increase of the RF losses. Lastly, sometimes, partial coating with low $\delta_0$ materials like graphite is used if the critical places for the development of MP discharges are restricted to small regions of the cavity geometry, and the added RF losses are affordable in terms of RF power and local heating of the cavity [2].

Owing to the high RF losses of the low $\delta_0$ this method is forbidden in superconducting cavities.

## 4.2 Avoid spatial focusing and resonant build-up

This method is the 'silver bullet' to kill most of the multipacting discharges. If we succeed to move the landing zone of secondary electrons to a region different from the take-off region, we have a fair chance that either the synchronization is lost or the impact energy goes out from the secondary multiplication range 100–1500 eV; or both.

This was the solution adopted to master the MP problem in superconducting cavities after the pioneering work of the Genoa Superconducting Radiofrequency Group in the late 1970s [7].

The desired effect was obtained by adopting a rounded shape in the equator region of the cavity equator like the cavity shown on Fig. 6 (spherical cells).

With that shape the electrons emitted at any place on the cavity surface always experience a force due to the strong component of the axial field along the axis $E_z$, sweeping the re-emitted electron away from the emission point, towards the cavity equator; furthermore the energy gain is usually higher than 1500 eV and the multiplication per impact is low.

At the equator the total electric field is null for symmetry reasons, the energy gain for the re-emitted electron is very low, usually lower than 100 eV, and again the secondary multiplication is lower than one.

In this way the re-emitted current decreases exponentially in a few hits, and the MP discharge is stopped.

## 5 Conclusions

The conditions leading to multipacting discharges in RF devices and accelerating structure were discussed.

In the case of axial symmetric structures some analytical approximate formulas, discussed in Section 3 are quite useful for a preliminary estimation of the RF field level leading to possible resonant discharges.

For arbitrary geometries, but also in axial symmetric structure, the only possible approach to evaluate the possibility of MP is the integration of the equation of motion of the emitted electron in the RF field and the thorough implementation of the secondary emission process at the cavity surface impact.

A very exhaustive discussion of the features of all the existing MP tracking codes for RF structures, either 2D or 3D, is reported in Ref. [8].